\documentstyle[preprint,aps]{revtex}
\begin{document}
\draft
\newfont{\form}{cmss10}
\newcommand{\unity}{1\kern-.65mm \mbox{\form l}}
\newcommand{\k}{\mbox{\form l}\kern-.6mm \mbox{\form K}}
\newcommand{\D}{D \raise0.5mm\hbox{\kern-2.0mm /}}
\newcommand{\A}{A \raise0.5mm\hbox{\kern-1.8mm /}}
\def\pmb#1{\leavevmode\setbox0=\hbox{$#1$}\kern-.025em\copy0\kern-\wd0
\kern-.05em\copy0\kern-\wd0\kern-.025em\raise.0433em\box0}

\def\D{\hbox{\hbox{${D}$}}\kern-1.9mm{\hbox{${/}$}}}
\def\kbar{\hbox{$k$}\kern-0.2truecm\hbox{$/$}}
\def\nbar{\hbox{$n$}\kern-0.23truecm\hbox{$/$}}
\def\pbar{\hbox{$p$}\kern-0.18truecm\hbox{$/$}}
\def\nhbar{\hbox{$\hat n$}\kern-0.23truecm\hbox{$/$}}

\title{Renormalization in light--cone gauge: how to do it in a consistent
way.}
\author{A. Bassetto}
\address{Dipartimento di Fisica ``G.Galilei", Via Marzolo 8 -- 35131
Padova, Italy \\ INFN, Sezione di Padova, Italy}
\maketitle
\vskip 4.0truecm
\begin{abstract}
We summarize several basic features concerning canonical equal time 
quantization and renormalization of  Yang--Mills theories in light--cone 
gauge. We describe a ``two component" formulation which is
reminiscent of the light--cone hamiltonian perturbation rules.
Finally we review the derivation of the one--loop Altarelli--Parisi
densities, using the correct causal
prescription on the ``spurious" pole.
\end{abstract}
\vskip 4.0truecm
Invited report at the Workshop ``QCD and QED in Higher Order", 
Rheinsberg, April 1996.
\vfill\eject

\narrowtext

\section{Introduction}
\label{prima}
Axial type gauges, characterized by the homogeneous $n^\mu A_\mu =0$ or 
inhomogeneous $n^\mu A_\mu = \phi$ conditions, $n^\mu$ being a fixed 
constant vector and $\phi$ a free field, have been considered long time
ago, in particular since the beginning of perturbative QCD calculations 
\cite{DDT}.

They are often called ``physical" or ``unitary" gauges, although this is 
not completely true, as it will appear in the sequel. Certainly, they trade 
manifest Lorentz covariance in favour of the absence of unphysical degrees
of freedom, at least in the homogeneous case \cite{BNS}.
For this reason they are particularly suitable in perturbative 
calculations: planar diagrams dominate in deep--inelastic scattering and are 
endowed with a transparent partonic interpretation.

In supersymmetric theories, the light--cone gauge ($n^2=0$) enjoys 
the property of having equal number 
of ``transverse" independent fields and of ``physical" excitations.
Finiteness of SUSY $N=4$ can thereby most naturally be proven
\cite{BDA}.

A further simplification occurs owing to the
decoupling of the Faddeev--Popov determinant, at least for trivial topological 
configurations.

Still one has to bear in mind that they are ``singular" gauges: delicate 
prescriptions are in order when handling Feynman propagators in 
perturbative calculations. Even more delicate is the issue concerning the 
possibility of regularizing and eventually renormalizing Green's functions.

This is the main topic discussed in the sequel: we shall use dimensional 
regularization throughout. The goal of bringing algebraic non covariant 
gauges to a level of accuracy comparable to the one obtained in the more 
familiar Feynman gauge has been achieved \cite{BNS} and is now a matter for 
textbooks.

Lorentz covariance is recovered in these gauges by the combined use of the 
Dirac formulation of constrained systems together with a weak condition, 
when necessary, to single the ``physical" Hilbert space out of an 
indefinite metric Fock space.
Lorentz covariance is achieved once all observable quantities possess
correct transformation properties under the Poincar\'e algebra, possibly 
restricted to the ``physical" subspace. This is exactly what the 
equivalence principle requires and has been carefully discussed in ref. 
\cite{BNS}.

Two gauge choices showed up to be viable so far, although on a quite 
different status: the spacelike planar gauge $n^\mu A_\mu = \phi, 
n^2<0$ and the light--cone gauge $n^\mu A_\mu =0, n^2 =0$. They share the 
following form of the free Feynman propagator

\begin{equation}
\label{uno}
D_{\mu\nu} (k) = {i \over k^2 + i \in} [-g_{\mu\nu} + {n_\mu k_\nu + n_\nu 
k_\mu \over nk} ]. 
\end{equation}

The quantity $(nk)^{-1}$ needs a prescription in order to represent a 
well defined distribution. We shall first briefly comment the
spacelike case and then focus our attention on the light--cone gauge.

\section{The spacelike option}
\label{seconda}
When $n^2<0$, one can choose $n_\mu = (0,0,0,1)$ without loss of 
generality. Then the singularity $(n k)^{-1}$ does not interfere with 
the causal Feynman poles at $k^2 =0$; in particular the integration 
contour can be Wick rotated without extra terms.

Canonical quantization suggests the Cauchy principal value (P) 
for $(n k)^{-1}$ in this case \cite{BLS}.

The field $\phi$ has the wrong sign in its quantum algebra, namely it 
behaves like a ``ghost".
Nevertheless, being a free field, it can be consistently excluded from the 
``physical" Hilbert space by means of the weak condition $\phi^{(-)} | 
\Phi_{\rm phys} > =0$. 

However ambiguities arise in higher orders: the only mathematically sound 
way to interpret $(nk)^{-2}$ is in the distribution sense

\begin{equation}
\label{due}
(nk)^{-2} \equiv - {d \over d(nk)} P\big({1 \over {nk}}\big), 
\end{equation}

which spoils positivity; as a consequence consistency with unitarity is 
not granted. The algebraic splitting formula 

\begin{eqnarray}
\label{tre}
P\big({1 \over {nk}}\big) P\big({1 \over {n(p-k)}}\big) &=&{1 \over 
{np}} \big[P\big({1 \over {nk}}\big)+\nonumber\\
&+& P\big( {1 \over {n(p-k)}}\big)\big] 
\end{eqnarray}

does not hold . As a matter of fact of fact the Poincar\'e-Bertrand 
term

\begin{equation}
\label{quattro}
- \pi^2 \delta (nk) \delta (np) 
\end{equation}

should be added, which has always been disregarded in practical 
calculations.  In order to justify this procedure, loop integrals 
require 
the use of peculiar functional spaces (Besov spaces) as well as delicate 
considerations concerning the adiabatic switching of the interaction.
This has been discussed at length in \cite{BSR}, where exponentiation of 
the Wilson loop up to the order $g^4$ has been proven as a test of 
gauge invariance. However, beyond this perturbative order, there is no 
guarantee of consistency; the renormalization proposed in \cite{MFA} has 
thereby to be regarded only in a formal sense. 

Finally the limit $n^2 
\rightarrow 0$ turns out to be singular and generally out of 
control when setting up renormalization.

\section{The light--cone choice}
\label{terza}

For all the previous reasons it is worth considering the light--cone 
gauge $n^u A_\mu =0, n^2 =0$, to be imposed in a strong sense, i.e. by means
of a Lagrange multiplier $\lambda$. It is not restrictive to choose $n_\mu 
={1\over{\sqrt 2}}(1,0,0,1)$. One easily recovers the expression 
(\ref{uno}) for the 
free Feynman propagator, but now  the singularity at $nk=0$ can interfere 
with the Feynman poles at $k^2=0$.

If P--prescription (or a sharp infrared cutoff) is adopted in analogy 
with the spacelike case, causality (and thereby analyticity) is violated. 
As a matter of fact the Cauchy principal value distribution is always
the sum 
of a causal pole and of an anti--causal one. The latter produces an extra 
unwanted term under Wick rotation through a pinch of the integration contour.

Power counting control of superficially divergent Feynman diagrams is lost 
together with all standard theorems (Weinberg -- BPHZ) which stand at the 
very basis of renormalization  \cite{CJS}.

A mismatch occurs between ultraviolet and collinear singularities; 
renormalization {\it constants} turn out to be momentum dependent.
In these conditions, although correct results for higher order 
contributions in particular instances cannot be excluded a priori if clever
recipes are followed, they are not supported by any sound general procedure.

Equal time canonical quantization induces a causal behaviour on the 
singularity $nk=0$ \cite{BDLS}:

\begin{eqnarray}
\label{cinque}
{1 \over k_0 - k_3} & \equiv & {1 \over {k_0 - k_3 + i \in{\rm sign} 
 (k_0 + k_3)} }= \nonumber\\ 
   &=& {k_0 + k_3 \over k^2_0 - k^2_3 + i \in } 
\end{eqnarray}

which, in turn, allows a Wick rotation without extra terms.

The first form of eq.(\ref{cinque}) was heuristically proposed by Mandelstam 
\cite{MDL}, the
second one  by Leibbrandt \cite{LBB} (ML prescription).

The free propagator now possesses two absorptive parts \cite{BAT}
\begin{eqnarray}
\label{sei}
&&{\rm disc}D_{\mu\nu}(k)= 2 \pi  \theta (k_0) \delta (k^2)\cdot 
\nonumber\\
&\cdot&\big[-g_{\mu\nu} + {n_\mu k_\nu  + n_\nu k_\mu \over k^2_\bot } 
{2 \hat n k \over n \hat n }\big]
- 2 \pi  \theta (k_0) \cdot \nonumber\\
&\cdot&\delta (k^2 + k^2_\bot)
{2 \hat n k \over n \hat n } {n_\mu k_\nu + n_\nu k_\mu \over 
k^2_\bot }, 
\end{eqnarray}

where $\hat n_\mu = {1 \over \sqrt 2}(1,0,0, -1) $ .

The second contribution has the wrong sign, namely it is of a ``ghost" 
type. We stress that it is not an optional choice, it is an 
unavoidable consequence of equal time canonical quantization. Its presence 
naturally protects the collinear behaviour $(k_\bot =0)$ of the propagator.

Negative norm states occur in the perturbative Fock space; however 
they are consistently expunged from the ``physical" Hilbert space by
imposing Gauss' law in a ``weak" sense \cite{BDLS}. In 
this  Hilbert space unitarity is automatically restored.

The possibility of a Wick rotation without extra terms leads to 
power counting control of superficially divergent graphs. Standard theorems 
are recovered, provided two separate countings are performed with respect to
a dilatation of all momentum components and of only ``transverse" ones. 
Convergence requires both indices to be negative.

Ultraviolet and infrared singularities become fully disentangled: Green 
functions in euclidean regions of momenta exhibit only ultraviolet 
singularities which appear as poles at $D=4$, $D$ being the number of 
dimensions.

One particle irreducible vertices turn out to have poles at $D=4$ with 
residues which sometimes involve non polynomialities with respect to 
``external" momenta of the type $(np)^{-1}$. Non local counterterms are 
thereby required, although of a very special kind \cite{BDS}.

After a careful study of all possible tensorial structures, after imposing Ward
identities, which are simple in light--cone gauge, and further technical 
conditions needed to match with the spacelike case, in ref. \cite{BDS} it 
has been shown that there is only one non local acceptable structure

\begin{equation}
\label{sette}
\Omega = (nD)^{-1} {n^\mu F_{\mu\nu}\hat n^\nu \over n \hat n},  
\end{equation}

$F_{\mu\nu}$ being the usual field tensor and $D$ the covariant 
derivative acting on the adjoint representation; $(nD)^{-1}$ is to 
be understood in a perturbative sense, with causal boundary conditions.

$\Omega$ is a covariant quantity with mass dimension equal to unity. It 
gives rise to the counterterm in the effective action 

\begin{equation}
\label{otto}
\Delta= \Omega n^\mu [D^\nu F_{\mu\nu} - g \bar \psi \tau 
\gamma_\mu \psi], 
\end{equation}

where one recognizes the classical equation of motion, as expected on 
general grounds \cite{VTBL}.

The canonical transformation

\begin{eqnarray}
\label{nove}
A_\mu^{(0)} &=& Z_3^{1/2} [A_\mu - (1 - \tilde Z_3^{-1}) n_\mu \Omega], 
\nonumber\\
\psi^{(0)} &=& Z_2^{1/2} ({\nhbar \nbar \over 2 n \hat n} + \tilde Z_2 
{\nbar 
\nhbar \over 2 n \hat n}) \psi, \nonumber\\
g_0 &=& Z_3^{-1/2} g, \nonumber\\
\lambda^{(0)}&=& Z_3^{-1/2}\lambda
\end{eqnarray}

relates bare and renormalized fields through the appearance of four 
renormalization constants. Only two of them $(Z_2, Z_3)$ are however 
independent, as it will be explained in the sequel.

All Green functions have been explicitly computed at one loop, in 
particular the renormalization constants at ${\cal O}(g^2)$ \cite{BNS}.
Results at two loop level have also been obtained. The correct 
exponentiation of a Wilson loop with light--like sides has been checked 
${\cal O}(g^4)$ together with a calculation  of the related 
anomalous dimensions at the same order 
\cite{BKN}.

One should also mention an interesting result concerning composite 
operators: it has been shown at any order in the loop expansion \cite{ABT}
that gauge invariant composite operators in light--cone gauge mix under 
renormalization only among themselves, at variance with their behaviour in 
covariant gauges \cite{JLK}.

\section{The two--component formulation}
\label{quarta}
We would like to discuss a ``two
component" formulation which may be useful in particular instances. 
Let's start from
the Green function generating functional

\begin{equation} 
\label{dieci}
W[J, \eta] = \int d[A_\mu, \lambda, \bar \psi, \psi] e^{i\int d^4x [{\cal 
L} + {\cal L}_s]}, 
\end{equation}

where
\begin{eqnarray}
\label{undici}
&&{\cal L}= - {1\over 4} F_{\mu\nu} F^{\mu\nu} + \lambda n A + \bar \psi (i
\D - m) \psi, \nonumber\\
&&{\cal L}_s= J_\mu A^\mu + \bar \eta \psi + \bar \psi \eta. 
\end{eqnarray}

$J$ and $\eta$ are external sources, $\lambda$ a Lagrange multiplier 
enforcing the condition $nA = A_- =0;$ colour indices are understood.
Let us also introduce the projection operators

\begin{equation}
\label{dodici}
P_+ = {\nhbar \nbar \over 2 n \hat n},\,\, P_- = {\nbar \nhbar
\over 2 n \hat n}, 
\,\,P_+ + P_- =1. 
\end{equation}

In light--cone gauge $W$ is gaussian with respect to the variables $A_+$ 
and $\chi$:

\begin{equation}
\label{tredici}
\chi = P_- \psi,\,\, \varphi = P_+ \psi,\,\, \psi = \varphi + \chi\,\,. 
\end{equation}

Similarly we define

\begin{equation}
\label{quattordici}
\xi = P_+ \eta,\,\, \zeta = P_- \eta .
\end{equation}

Then, integrating over $A_+$ and $\chi$, we get

\begin{eqnarray}
\label{quindici}
&&W={\rm exp} [{i \over 2}\int\!\big( J^+ \partial^{-2}_- J^+ + \bar \xi {i 
\gamma^+ \partial_+ \over \partial_+ \partial_-} \xi) d^4x ]\cdot \nonumber\\
&&\cdot \int d[A_\alpha, \varphi, \bar \varphi] e^{i \int\big( {\cal L}_{eff} 
+ {\cal L}_{mix} + \hat {\cal L}_s\big) d^4x} ,
\end{eqnarray}

where 
\begin{eqnarray}
\label{sedici}
{\cal L}_{eff}&=&- {1\over 4} F_{\alpha\beta} F_{\alpha\beta} + \partial_+ 
A_\alpha \partial_- A_\alpha+ \nonumber\\
&+& i \bar\varphi \gamma^+ \partial_+ \varphi - 
{1\over 2} K^2 + {1\over2} \bar \varphi (i\gamma^\alpha {\cal D}_\alpha - 
\nonumber\\
&-& m){i\gamma^+ \partial_+ \over \partial_+\partial_-}
(i \gamma^\alpha {\cal D}_\alpha -m) 
\varphi, \nonumber\\
{\cal L}_{mix}&=&-K \partial_-^{-1} J^+ + {1\over 2} \bar\xi {i\gamma^+ 
\partial_+ \over \partial_+\partial_-} (i\gamma^\alpha {\cal D}_\alpha- 
\nonumber\\
&-&m)\varphi + {1\over 2} \bar\varphi (i \gamma^\alpha {\cal D}_\alpha - m) 
{i \gamma^+\partial_+ \over \partial_+ \partial_-}\xi , \nonumber\\
\hat{\cal L}_s &=& J^\alpha A_\alpha + \bar \zeta \varphi + \bar\varphi 
\zeta.
\end{eqnarray}

In eqs.(\ref{sedici}) $\alpha =1,2$ and

\begin{equation}
\label{diciassette}
K= \partial^{-1}_- [D_\alpha \partial_- A_\alpha + g \bar \varphi \gamma^+ T 
\varphi].
\end{equation}

Moreover $\partial^{-1}_-$ and $(\partial_+ \partial_-)^{-1}$ have always to be 
understood with causal boundary conditions.

Only ``transverse" fields, $A_\alpha$ and $\varphi$, appear in 
eq.(\ref{quindici}); the
dependent fields $A_+$ and $\chi$ can be expressed in terms of $A_\alpha$ 
and $\varphi$, although in a non local way. Their Green functions can 
also be expressed by means of ``bridge identities" (BI) \cite{BAT} in terms of 
the independent ``transverse" ones: in particular the renormalization
constants $\tilde Z_2$ and $\tilde Z_3$ can be 
obtained as dependent quantities at any order in the loop expansion.

The (BI) read

\begin{eqnarray}
&&\partial_-^2{\delta \over i\delta J^+}=J^++D_{\alpha}[{\delta \over
i \delta J_{\alpha}}]\partial_-{\delta \over i \delta 
J_{\alpha}}+\nonumber\\
&&+g {\delta \over i\delta \zeta}\gamma^+T{\delta \over i \delta \bar 
\zeta},\nonumber\\
&&2\partial_-{\delta \over i \delta \bar \xi}=i \gamma^+\xi+i \gamma^+
\cdot\nonumber\\
&&\cdot\big[i\gamma^{\alpha}{\cal D}_{\alpha}[{\delta\over i\delta 
J_{\alpha}}]-m\big]{\delta \over i\delta \bar \zeta},\nonumber\\
&&2\partial_-{\delta \over i \delta  \xi}=i \bar \xi \gamma^+ +
{\delta \over i\delta \zeta}\cdot\nonumber\\
&&\cdot\big[i\gamma^{\alpha}{\cal D}_{\alpha}[{\delta\over i\delta 
J_{\alpha}}]-m\big]i \gamma^+,\nonumber\\
\end{eqnarray}

where the operators are supposed to act on $W$ of eq.(\ref{quindici}).
These identities hold to any order in perturbation theory and usually 
mix terms with different powers of the coupling constant $g$.

If only transverse Green functions are sought, one can set $J^+$ and 
$\xi$ equal to zero in eq.(\ref{quindici}).

In the ``two component" formulation new vertices appear with non polynomial
character and complicate topology, already at the tree level.
They bear no simple  relation with the vertices of the ``four component" 
formulation. They are reminiscent of the vertices occurring in light--cone 
hamiltonian theory \cite{BLK}. However the ML prescription prevents from 
integrating first over the $(+)$--momentum components; 
a transition to the old--fashioned perturbation theory is 
thereby impossible, unless peculiar subtractions are performed 
``step--by--step" \cite{BAP}.

Renormalization cannot be directly proven in the ``two component" formulation,
because the basic theorems do not apply. However, from the transformation 
(\ref{nove}), one can easily obtain

\begin{equation}
\label{diciotto}
W [J_\alpha, \zeta]= \int d[A_\alpha, \varphi, \bar \varphi]
e^{i\int({\cal
L}_R + {\cal L}_s)d^4x}, 
\end{equation}

where
\begin{eqnarray}
\label{diciannove}
&{\cal L}_R&=-{Z_3 \over 4} F_{\alpha\beta} F_{\alpha\beta} + Z_3 
\partial_+ A_\alpha \partial_- A_\alpha+ \nonumber\\
&+&i Z_2 \bar\varphi \gamma^+ \partial_+ \varphi
- {Z_3 \over 2} \big(\partial_-^{-1} [D_\alpha 
\partial_- A_\alpha + \nonumber\\
&+&g {Z_2 
\over Z_3} \bar \varphi \gamma^+ T \varphi] \big)^2 +
{Z_2 \over 2} [\bar \varphi (i \gamma^\alpha {\cal D}_\alpha - m)\cdot 
\nonumber\\
&\cdot& {i\gamma^+ 
\partial_+ \over \partial_+\partial_-} (i\gamma^\alpha {\cal D}_\alpha - m) 
\varphi ].
\end{eqnarray}

``Unphysical" renormalization constants $\tilde Z_2$ and $\tilde Z_3$ no 
longer occur, nor the non local quantity $\Omega$. They are buried 
in the non local structures which are produced when developing 
perturbation theory starting from the functional (\ref{diciotto}). 

\section{The Altarelli--Parisi Densities}
\label{quinta}

One loop  Altarelli--Parisi (AP) splitting functions have been correctly 
recovered  in this causal light--cone  formulation \cite{BAS}; the basic 
new feature is the appearance of the well--defined $(1-x)^{-1}_+$ 
distribution already in the ``real" contributions.

Let us briefly review this derivation.

Kinematics can be usefully parametrized as

\begin{eqnarray}
\label{venti}
p_\mu&=&(P+{p^2\over 4P},\, \underline{0}\,, P-{p^2\over 4P}),\nonumber\\ 
k_\mu&=&(\xi P+{k^2+k^2_\bot\over 4\xi P},\underline{k}_\bot, \xi 
P-{k^2+k^2_\bot\over 4\xi P}), 
\end{eqnarray}

and

\begin{equation}
\label{ventuno}
n_\mu=({np\over 2P},\, \underline{0}\,, -{np\over 2P}), \,\,\,
\hat n_\mu=({P\over np},\, \underline{0}\,, {P\over np}).
\end{equation}

Here $\xi$ represents the fraction of the (large) longitudinal 
component $P$ of 
the incoming quark momentum $p$ (small $p^2< 0$), 
carried by $k$. 

The AP density is the coefficient of the term 
$\log\mid{Q^2\over p^2}\mid$, in the propagation kernel , when $p^2\to
0$ , $-Q^2$ being the (large) virtuality of the external current up to 
which the vector $k$ has to be integrated.
To the ``real" part of the kernel $K^{(a)}$ we associate the 
quantity \cite{CFP}

\begin{eqnarray}
\label{ventidue}
&K&^{(a)}_q(x,\mid{Q^2\over p^2}\mid)=g^2c_F\int{d^4k\over 
(2\pi)^4}\cdot \nonumber\\
&\cdot&\delta(1-{nk\over xnp}){1\over (k^2)^2}\cdot Tr[{\nbar \over 
4nk}\kbar\gamma^\mu\pbar\gamma^\nu\kbar]\cdot \nonumber\\
&\cdot& disc[D_{\mu\nu}(p-k)],
\end{eqnarray}

the discontinuity being the one of eq.(\ref{sei}). The spin trace is 
self--explanatory but the factor ${\nbar\over 4nk}$, which is introduced to 
project the ``leading-log" contribution; $c_F$ is the usual colour factor. 
We can safely work in four dimensions, as no ultraviolet (UV)
singularities occur since 
we are evaluating an absorptive part and no infrared (IR)
singularities either, as 
long as $p^2<0$, thanks to the ML prescription.
A straightforward calculation gives

\begin{eqnarray}
\label{ventitre}
&&K^{(a)}_q={g^2 c_F\over 8\pi^2}\int^{-Q^2}_{-p^2}{d\mid k^2\mid\over 
\mid k^2\mid}\int d(k_\bot^2)\cdot\nonumber\\
&\cdot&\delta(1-x-{k^2_\bot\over \mid k^2\mid})
[{1-x\over 
\mid k^2\mid}+{2x\over k^2_\bot}]-
\nonumber\\
&&-{g^2c_F\over 8\pi^2}\int^{-Q^2}_{-p^2}{d\mid k^2\mid\over \mid 
k^2\mid}\int d(k^2_\bot){2\over k^2_\bot}(1-{k^2_\bot\over \mid 
k^2\mid})\cdot \nonumber\\
&\cdot&\theta(\mid k^2\mid-k^2_\bot)
\delta\left((1-x)(1-{k^2_\bot\over \mid k^2\mid})\right),
\end{eqnarray}

the second addendum arising from the presence of the ghost. Both 
contributions are singular at $x=1$, but they nicely combine; we
have indeed 

\begin{eqnarray}
\label{ventiquattro}
&&K^{(a)}_q={g^2 c_F\over 8\pi^2}\log\mid{Q^2\over 
p^2}\mid\left[(1-x+{2x\over 1-x})-\right. \nonumber\\
&&\left.-2\delta(1-x)\int_0^1{dy\over 
1-y}\right],
\end{eqnarray}

namely the well defined distribution
\begin{equation}
\label{venticinque}
K_q^{(a)}={g^2 c_F\over 8\pi^2}\log\mid{Q^2\over p^2}\mid[-1-x+{2\over 
(1-x)_+}]. 
\end{equation}

We notice that the IR singularity at $x=1$ is fully regularized by the 
ghost, already in the diagram describing the ``real" contribution.

The one--loop self--energy, regularized in $D=2w$
dimensions, has been discussed at length in 
ref. \cite{BNS}. It has 
the expression ($\mu$ is here the renormalization scale)

\begin{eqnarray}
\label{ventisei}
&\Sigma&(p)=-{ig^2 c_F\over 16\pi^2}({-p^2\over 
4\pi\mu^2})^{w-2}\left[-{\pbar \over \sin (\pi w)}\right.\cdot \nonumber\\
&\cdot&{\pi\Gamma(w)\over 
\Gamma(2w-2)}+\left.2{\nbar\hat np\over n\hat n}[(1+{\Gamma(w)\Gamma(w-1)\over 
\Gamma(2w-2)})\cdot\right.\nonumber\\
&\cdot&\left. \Gamma(1-w)+ {\pi^2\over 6}+\psi^\prime(2)]\right].
\end{eqnarray}

It exhibits the nice feature of having no IR singularities as long as 
$p^2<0$, at variance with expressions obtained in previous treatments, in 
which P-prescription was adopted. 

From eq.(\ref{ventisei}) one easily realizes that the one loop radiative 
correction at the pole $p^2=0$  of the fermion propagator
renormalized in the minimal subtraction scheme is

\begin{equation}
\label{ventisette}
\Delta S^R=-{3g^2c_F\over 16\pi^2}\log({-p^2\over 4\pi\mu^2})+f.t.
\end{equation}

where $f.t.$ refers to terms which are finite in the limit $p^2\to 0$.

This result, together with eq.(\ref{venticinque}), finally gives

\begin{eqnarray}
\label{ventotto}
&&K_q(x,p^2)={g^2 c_F\over 8\pi^2}\log\mid{p^2\over \mu^2}\mid[1+x-
\nonumber\\
&&-{2\over (1-x)_+}-
{3\over 2}\delta(1-x)]+f.t.
\end{eqnarray}

and one recognizes the flavour non singlet AP density

\begin{eqnarray}
\label{ventinove}
&&P_q^q(x)={\alpha_s c_F\over 2\pi}[-1-x+{2\over (1-x)_+}+ \nonumber\\
&+&{3\over 
2}\delta(1-x)]\equiv{\alpha_s c_F\over 2\pi}\big({1+x^2\over 
1-x}\big)_+,
\end{eqnarray}

with $\alpha_s=g^2/4\pi$.

We notice that, were we interested 
in computing branching probabilities, both the ghost and the virtual radiative
corrections at the fermion pole should be omitted, and the IR singularity 
at $x=1$ would be fully exposed 

\begin{equation}
\label{trenta}
\hat P_q^q={\alpha_s c_F\over 2\pi}{1+x^2\over 1-x}.
\end{equation}

Should we instead be interested in Sudakov form factor, the gluon
radiation (but not the ghost one!) should be inhibited in the
absorptive part and the usual result would be easily recovered.

One-loop unitarity sum rules relate real $r(x)$ and virtual $v(x)$
contributions, as is well known \cite{BCM}. In our approach both
quantities are separately well defined, as anytime a gluon is summed
over, the ghost is standing by it \cite{BAT}, 
to protect its IR behaviour with
the appropriate $\delta-$measure.

As a matter of fact, in the flavour non-singlet case, we have

\begin{eqnarray}
\label{trentuno}
&&v_q^q(x)=-{1\over 2}\delta(1-x)\int_0^1 dy 
[r_q^q(y)+r_q^g(y)]= \nonumber\\
&&=-{\alpha_sc_F \over 4\pi}\delta(1-x)\int_0^1 dy
\left([-1-y+\right.\nonumber\\
&&+\left.{2\over(1-y)_+}]
+[-2+y+{2\over y_+}]\right)= \nonumber\\
&&={3\alpha_sc_F \over 4\pi}\delta(1-x),
\end{eqnarray}

and thereby

\begin{equation}
\label{trentadue}
P_q^q(x)=r_q^q(x)+v_q^q(x),
\end{equation}

as expected. We stress that $v_q^q(x)$ is positive, at variance with 
previous treatments, owing to the ghost contribution. In turn the 
real contribution is negative due to its ``overshielding". In spite
of those paradoxical behaviours, they nicely combine to give the
correct answer for any quantity of physical interest.

Now we repeat the calculation for the
gluon--gluon case. 

We introduce the vectors
\begin{equation}
\label{trentaquattro}
e_\mu^{(\beta)}(k)=-g_{\mu\beta}+{n_\mu k_\beta\over [nk]},\qquad \beta=1,2 .
\end{equation}

These vectors enjoy the property of being orthogonal to both $n_\mu$ and 
$k_\mu$. We have indeed 

\begin{equation}
\label{trentacinque}
n^\mu e_\mu^{(\alpha)}=k^\mu e_\mu^{(\alpha)}=0.
\end{equation}

When $k^2=0$, $nk={k^2_\bot\over 2\hat n_+k_-}$ and $e^{(\alpha)}_\mu$ become 
the two (physical) polarization vectors.

If we are interested in structure function, 
the vector $q=p-k$ is on--shell as we are 
computing just an absorptive part, the vector $p$ is slightly off-shell 
and
the vector $k$ is spacelike. As $\xi$  cannot vanish
in the kinematical region of interest, the prescription 
in eq.(\ref{cinque}) is irrelevant both for the vectors 
$e^{(\alpha)}_\mu(p)$ and $e^{(\beta)}_\rho(k)$ .
One can also show that

\begin{eqnarray}
\label{trentasei}
&&\sum^2_{\alpha=1}e^{(\alpha)}_\mu(k) 
e_\nu^{(\alpha)}(k)=-d_{\mu\rho}(k)d^\rho_\nu(k)=\nonumber\\
&&=
d_{\mu\nu}(k)-
{n_\mu n_\nu k^2 \over [nk]^2}.
\end{eqnarray}

Then we define the tensor

\begin{eqnarray}
\label{trentasette}
&&T^{\nu\nu'}={1\over 
2}\sum_{\alpha, \beta}\left[e^{(\alpha)}_\mu(p)e^{(\beta)}_\rho(k)
V^{\mu\rho\nu}\right]\nonumber\\
&\cdot&\left[e^{(\alpha)}_{\mu'}(p)e^{(\beta)}_{\rho'}(k)V^{\mu'\rho'\nu'}
\right],
\end{eqnarray}

$V^{\mu\rho\nu}$ being the triple gluon vertex
(we have here averaged over initial polarizations). 

The usual definition of the gluon--gluon kernel
entails the quantity

\begin{eqnarray}
\label{trentotto}
&&K^{(a)}_g=ig^2 c_A \int {d^4k\over (2\pi)^4}\delta\big(1-{nk\over 
xnp}\big)\cdot \nonumber\\
&\cdot&{T^{\mu\nu}\over (k^2)^2}disc[D_{\mu\nu}(p-k)],
\end{eqnarray}

$c_A$ being the relevant colour factor.

A lengthy but straightforward calculation gives for $x>0$

\begin{eqnarray}
\label{trentanove}
&&K^{(a)}_g={g^2 c_A\over 4\pi^2}\log\mid{Q^2\over 
p^2}\mid\Big[x(1-x)+\nonumber\\
&+&{1-x\over x}+{x\over 1-x}-\delta(1-x)\int^1_0{dy\over 
1-y}\Big]\equiv \nonumber\\
&\equiv&{g^2 c_A\over 4\pi^2}\log\mid{Q^2\over 
p^2}\mid\cdot\Big[x(1-x)+\nonumber\\
&+&{1-x\over x}-1+{1\over (1-x)_+}\Big].
\end{eqnarray}

The ghost is responsible for the term with the $\delta$--function. Again both 
contributions are singular at $x=1$, but the IR singularity at $x=1$ is 
exactly regularized already in this real kernel.

The one loop expression of the self--energy tensor has been 
completely evaluated \cite{MDAL}. It will not be reported here. We
give instead the one loop radiative corrections to the transverse 
components of the vector propagator for pure 
Yang--Mills theory, 
renormalized in the minimal subtraction scheme 

\begin{equation}
\label{quaranta}
\Delta D^R_{\alpha\beta}=-{ig_{\alpha\beta}\over p^2}{g^2 c_A\over 
16\pi^2}{11\over 3}\log({4\pi\mu^2\over -p^2})+f.t.
\end{equation}

where f.t. refer again to terms which are finite in the limit 
$p^2\rightarrow 0$.

Collecting the one--loop virtual radiative corrections at the gluon pole 
with the expression (\ref{trentanove}) we get for $x>0$

\begin{eqnarray}
\label{quarantuno}
&&K_g={g^2 c_A\over 4\pi^2}\log\big({4\pi\mu^2\over -p^2}\big)[x(1-x)+{1-x\over 
x}-\nonumber\\
&-&1+{1\over (1-x)_+}+{11\over 12}\delta(1-x)]+f.t.,
\end{eqnarray}

leading to the corresponding well known AP density

\begin{eqnarray}
\label{quarantadue}
&&P^g_g={\alpha_s\over 2\pi}2c_A\left[{1-x\over x}+\big({x\over 1-x}\big)_+ 
+\right.\nonumber\\
&+&\left.x(1-x)-{1\over
12}\delta(1-x)\right].
\end{eqnarray}

To (\ref{quarantadue}) one should 
add the quark contribution we have disregarded in 
(\ref{quaranta}), giving the extra term 
$-{\alpha_s\over 6\pi}n_F\delta(1-x)$
($n_F$ being the 
flavour number); it does not entail any difference with respect
to previous treatments.
Again, when computing the probability density for real gluon emission, 
ghost contribution and virtual corrections should be omitted, thereby 
recovering full symmetry under the exchange $x\leftrightarrow 1-~x~.$

Finally, in the gluon case we can check again the unitarity sum rule

\begin{eqnarray}
\label{quarantatre}
&&v_g^g(x)=-{1\over 2}\delta(1-x)\int_0^1 dy 
[r_g^g(y)+2r_g^q(y)]=\nonumber\\
&=&-{\alpha_s \over 2\pi}\delta(1-x)\int_0^1 dy\left(c_A[y(1-y)
-2+\right.\nonumber\\
&+&\left.{1\over (1-y)_+}+{1\over y_+}]+{n_F \over 
2}[y^2+(1-y)^2]\right)=\nonumber\\
&=&{\alpha_s \over 2\pi}\delta(1-x)[{11\over 6}c_A-{1\over 3}n_F]. 
\end{eqnarray}

As a concluding remark, we hope that the successful calculation we
have just reported may encourage people to
apply the 
procedure we have described above, in a systematic way
to higher order calculations.

\end{document}